\documentclass{SciPost}

\hypersetup{
    colorlinks,
    linkcolor={red!50!black},
    citecolor={blue!50!black},
    urlcolor={blue!80!black}
}

\usepackage[bitstream-charter]{mathdesign}
\usepackage{young}
\usepackage{tikz}
\usepackage{tkz-euclide}
\DeclareSymbolFont{usualmathcal}{OMS}{cmsy}{m}{n}
\DeclareSymbolFontAlphabet{\mathcal}{usualmathcal}

\fancypagestyle{SPstyle}{
\fancyhf{}
\lhead{\colorbox{scipostblue}{\bf \color{white} ~SciPost Physics Proceedings }}
\rhead{{\bf \color{scipostdeepblue} ~Submission }}

\fancyfoot[C]{\textbf{\thepage}}
}
\begin{document}
%
\def\qdots{\mathinner{\mkern1mu\raise1pt\vbox{\kern7pt\hbox{.}}\mkern2mu \raise4pt\hbox{.}\mkern2mu\raise7pt\hbox{.}\mkern1mu}}
\newcommand{\hash}{\#} 
\def\ch{\mathop{\rm char}\nolimits}
\def\Z{{\mathbb Z}}
\def\N{{\mathbb N}}
\def\R{{\mathbb R}}
\def\C{{\mathbb C}}
\def\gl{\mathfrak{gl}}
\def\g{\mathfrak{g}}
\def\h{\mathfrak{h}}
\def\so{\mathfrak{so}}
\def\sp{\mathfrak{sp}}
\def\osp{\mathfrak{osp}}
\def\pso{\mathfrak{pso}}
\renewcommand{\theequation}{\arabic{section}.{\arabic{equation}}}
\allowdisplaybreaks 

\pagestyle{SPstyle}

\begin{center}{\Large \textbf{\color{scipostdeepblue}{
Spectrum of the parity operator for parafermions and parabosons from character formulas\\
}}}\end{center}

\begin{center}\textbf{
Neli I. Stoilova\textsuperscript{1$\star$} and
Joris Van der Jeugt\textsuperscript{2$\dagger$}
}\end{center}

\begin{center}
{\bf 1} Institute for Nuclear Research and Nuclear Energy, Bulgarian Academy of Sciencies, 
Boul.\ Tsarigradsko Chaussee 72, 1784 Sofia, Bulgaria
\\
{\bf 2} Department of Mathematics, Computer Science and Statistics, Ghent University,
Krijgslaan 281-S9, B-9000 Gent, Belgium
\\[\baselineskip]
$\star$ \href{mailto:email1}{\small stoilova@inrne.bas.bg}\,,\quad
$\dagger$ \href{mailto:email2}{\small joris.vanderjeugt@ugent.be}
\end{center}

\definecolor{palegray}{gray}{0.95}
\begin{center}
\colorbox{palegray}{
  \begin{tabular}{ll}
 \hspace{-11pt} \begin{minipage}{0.18\textwidth}
    \includegraphics[width=28mm]{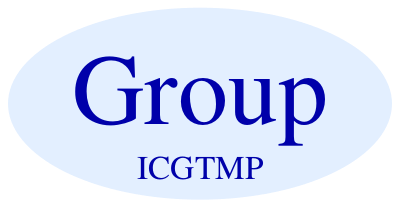}
  \end{minipage}
  &
  \begin{minipage}{0.75\textwidth}
    \begin{center}
    {\it 36th International Colloquium on Group Theoretical Methods in Physics}\\
    {\it Valladolid, 13-17 July 2026} \\
    \doi{10.21468/SciPostPhysProc.?}\\
    \end{center}
  \end{minipage}
\end{tabular}
}
\end{center}

\section*{\color{scipostdeepblue}{Abstract}}
\textbf{\boldmath{%
In a recent paper~\cite{SV2026} Green's triple relations for parafermions and parabosons were extended by including a new operator $P$, referred to as the parity operator.
The spectrum of $P$ in Fock space representations was computed, making use of the action of parafermion (resp.\ paraboson) creation and annihilation operators.
In the current paper, we adopt a different method to compute the spectrum of $P$, namely using character formulas.
For a set of $n$ parafermions, the action of $P$ is deduced from character formulas of $\so(2n+2)$.
For a set of $n$ parabosons, the known spectrum of $P$ yields new character formulas of particular $\osp(2|2n)$ representations.
}}

\vspace{\baselineskip}

\noindent\textcolor{white!90!black}{%
\fbox{\parbox{0.975\linewidth}{%
\textcolor{white!40!black}{\begin{tabular}{lr}%
  \begin{minipage}{0.6\textwidth}%
    {\small Copyright attribution to authors. \newline
    This work is a submission to SciPost Phys. Proc. \newline
    License information to appear upon publication. \newline
    Publication information to appear upon publication.}
  \end{minipage} & \begin{minipage}{0.4\textwidth}
    {\small Received Date \newline Accepted Date \newline Published Date}%
  \end{minipage}
\end{tabular}}
}}
}


\vspace{10pt}
\noindent\rule{\textwidth}{1pt}
\tableofcontents
\noindent\rule{\textwidth}{1pt}
\vspace{10pt}


\setcounter{equation}{0}
\section{Introduction}
\label{sec1}

For a standard set of $n$ bosons or fermions, the Fock space is particularly simple, and the number of particles in a basis state is apparent. 
Then the parity operator $P$ takes the value $+1$ (resp.\ $-1$) when this number is even (resp.\ odd).
For parabosons or parafermions, the basis states in a Fock space representation are very complicated~\cite{parafermion,paraboson}, hence at first sight it is not clear how to define a parity operator.
Following Green's approach, where creation and annihilation operators for a set of $n$ parafermions or parabosons are determined by triple relations~\cite{Green},
a parity operator $P$ was introduced by extending these triple relations in~\cite{SV2026}.
As a consequence, the algebraic structure determined by the $n$ creation and $n$ annihilation operators and their triple relations was also extended: 
for parafermions, this led to an extension from the Lie algebra $\so(2n+1)$ to $\so(2n+2)$;
for parabosons, this led to an extension from the Lie superalgebra $\osp(1|2n)$ to $\osp(2|2n)$.

The main interest is the determination of the action of $P$ in Fock spaces.
For parafermions, these Fock spaces are irreducible finite-dimensional unitary representations of $\so(2n+1)$, characterized by a positive integer $p$, and denoted by $W(p)$.
These Fock spaces can be extended in two ways to representations of $\so(2n+2)$, denoted by $W_+(p)$ and $W_-(p)$~\cite{SV2026}.
The main computational result of~\cite{SV2026} was the determination of the action of $P$ in a Gelfand-Zetlin (GZ) basis of these Fock spaces. 
In the current paper, we greatly simplify this computation by using special expressions for the characters of these $\so(2n+2)$ representations.

For parabosons, the Fock spaces are irreducible infinite-dimensional unitary lowest weight representations of $\osp(1|2n)$, also characterized by a positive integer $p$, and denoted by $V(p)$.
These Fock spaces are extended in two ways to representations of $\osp(2|2n)$, denoted by $V_+(p)$ and $V_-(p)$~\cite{SV2026}.
Also in this case, the action of $P$ in a GZ-basis of these Fock spaces was determined in~\cite{SV2026}, in a computationally hard way.
For $\osp(2|2n)$ however, character formulas of infinite-dimensional lowest weight representations are not known in general.
So here we can work the other way round: we use the action of $P$ determined in~\cite{SV2026} to construct the characters of $V_{\pm}(p)$, yielding new and interesting character formulas.

The new approach to the action of $P$ has some other advantages.
In particular, all expressions for eigenvalues of $P$ are highly simplified using the notation of partitions. 
In Section~\ref{sec2} we recall some standard notations related to a partition $\lambda$, and define $|\lambda|_a$, the less known alternating sum of parts of a partition $\lambda$.
Quite surprisingly, it is this alternating sum of parts that plays a crucial role in the eigenvalue expressions.
So we also give a number of properties of $|\lambda|_a$ in this section.
In Section~\ref{sec3} we recall the definition of the parafermion parity operator as given in~\cite{SV2026}. 
Using specific characters of $\so(2n+2)$, the action of $P$ in Fock spaces is determined in Section~\ref{sec4}.
In Section~\ref{sec5}, the paraboson parity operator is presented as defined in~\cite{SV2026}.
The action of $P$ in paraboson Fock spaces is then used to derive some character formulas of $\osp(2|2n)$ in Section~\ref{sec6}, and some spectrum generating functions are determined.
We end the paper by some conclusions in Section~\ref{sec7}.

\setcounter{equation}{0}
\section{Partitions and their alternating sum of parts}
\label{sec2}

In this paper, $n$ is a fixed positive integer, and all partitions have length less than or equal to~$n$.
We refer to Macdonald for notations and properties of partitions~\cite{Macdonald}.
Formally, a {\em partition} $\lambda$ is a sequence of non-negative integers $\lambda=(\lambda_1,\lambda_2,\ldots,\lambda_n)$ with $\lambda_1\geq \lambda_2\geq \cdots \geq\lambda_n\geq 0$.
The numbers $\lambda_i$ ($i=1,2,\ldots$) are the {\em parts} of the partition.
The {\em Young diagram} of $\lambda$ is a set of boxes arranged in left-aligned rows with row lengths $\lambda_1$, $\lambda_2$, $\ldots$.
For example, for $n=5$ and $\lambda=(3,2,2,1,0)$ the Young diagram is
\begin{equation}
\begin{Young}
&&\cr
&\cr
&\cr
\cr
\end{Young}
\end{equation}
A partition is often identified with its Young diagram, and it is natural to speak of the rows and columns of a Young diagram or of the partition.
The {\em weight} of a partition $\lambda$ is the sum of its parts and denoted by $|\lambda|$: $|\lambda|=\sum_{i=1}^n \lambda_i$.
The {\em length} of $\lambda$, denoted by $\ell(\lambda)$, is the number of nonzero parts of $\lambda$.
In the above example, $|\lambda|=8$ and $\ell(\lambda)=4$.

The {\em conjugate partition} of $\lambda$, denoted by $\lambda'$, is found by reflecting the Young diagram of $\lambda$ with respect to the main diagonal. 
In other words, the column lengths of the Young diagram of $\lambda$ are the parts of $\lambda'$.
For the above example, the reflected Young diagram is
\begin{equation}
\begin{Young}
&&&\cr
&&\cr
\cr
\end{Young}
\end{equation}
and hence $\lambda'=(4,3,1)=(4,3,1,0,0)$.

A notion that is less used in combinatorics is the {\em alternating sum of parts} of $\lambda$, denoted by $|\lambda|_a$:
\begin{equation}
|\lambda|_a = \lambda_1-\lambda_2+\lambda_3-\lambda_4+\cdots = \sum_{i=1}^n (-1)^{i+1} \lambda_i.
\end{equation}
In the above example, $|\lambda|_a=2$.
Let us consider some properties of $|\lambda|_a$.
From the Young diagram of a general partition it is clear that $\lambda_1-\lambda_2$ is equal to the number of columns of length 1,
$\lambda_3-\lambda_4$ is equal to the number of columns of length 3, etc.
Hence $\lambda_1-\lambda_2+\lambda_3-\lambda_4+\cdots$ is equal to the number of columns of odd length, see the scheme below.
\[
\begin{tikzpicture}[scale=0.6]
\def\lone{12}
\def\ltwo{9}
\def\lthree{6}
\def\lfour{3}
\foreach \y/\len in {0/\lone, 1/\ltwo, 2/\lthree, 3/\lfour} {
    \foreach \x in {0,...,\numexpr\len-1} {
        \draw (\x,-\y) rectangle (\x+1,-\y-1);
    }
}
\draw[decorate,decoration={brace,amplitude=6pt,mirror}]
    (\ltwo, -1.05) -- (\lone, -1.05)
    node[midway, below=8pt] {$\lambda_1-\lambda_2$};
\draw[decorate,decoration={brace,amplitude=6pt,mirror}]
    (\lfour, -3.05) -- (\lthree, -3.05)
    node[midway, below=8pt] {$\lambda_3-\lambda_4$};
\node at (1.5,-4.5) {$\vdots$};
\end{tikzpicture}
\]
In other words, $|\lambda|_a$ is equal to the number of odd parts in $\lambda'$:
\begin{equation}
|\lambda|_a = \hash(\hbox{odd parts }\lambda').
\label{oddparts}
\end{equation}
Then we also have
\begin{equation}
|\lambda'|_a = \hash(\hbox{odd parts }\lambda).
\end{equation}
We also need the following formula
\begin{equation}
|\lambda'|_a= \frac12 \Big(n-\sum_{i=1}^n (-1)^{\lambda_{i}}\Big).
\label{26}
\end{equation}
This is easy to verify, starting from the right hand side:
\begin{align*}
\frac12\Big(n-\sum_{i=1}^n (-1)^{\lambda_{i}}\Big)&=
\frac12\Big(n-\hash(\hbox{ even parts }\lambda) + \hash(\hbox{ odd parts }\lambda)\Big) \\
&=
\frac12\Big(2\hash(\hbox{ odd parts }\lambda)\Big) = \hash(\hbox{ odd parts }\lambda) = |\lambda'|_a,
\end{align*}
where we use the obvious fact that $n$ minus the number of even parts is equal to the number of odd parts.

The remaining notions that are needed here are more standard and appear in~\cite{Macdonald}.
If $\lambda$ and $\mu$ are two partitions and $\lambda_i \geq \mu_i$ for all $i$, then we write $\lambda\supset\mu$, 
and the Young diagram of $\theta=\lambda-\mu$ is called a {\em skew Young diagram}.
It consists of the Young diagram $\lambda$ in which the boxes of $\mu$ are removed.
Such a skew diagram is called a {\em horizontal strip} if it has at most one box in each column.
A necessary and sufficient condition for $\theta=\lambda-\mu$ to be a horizontal strip is that the sequences $\lambda$ and $\mu$ are {\em interlaced}, which means that
\[
\lambda_1\geq\mu_1\geq\lambda_2\geq\mu_2\geq\cdots.
\]

Partitions play a fundamental role in symmetric functions and representation theory. 
An important basis of the ring of symmetric functions over the integers is the set of Schur functions $s_\lambda$ with $\lambda$ a partition.
Recall that
\begin{equation}
s_\lambda(x_1,x_2,\ldots,x_n)= \frac{\det(x_i^{\lambda_j+n-j})_{1\leq i,j\leq n}}{\det(x_i^{n-j})_{1\leq i,j\leq n}}.
\end{equation}
In representation theory, the Schur function $s_\lambda(x_1,x_2,\ldots,x_n)$ is the character of the irreducible representation $L_\lambda$ of $\gl(n)$ with highest weight $\lambda=\lambda_1\epsilon_1+\cdots +\lambda_n\epsilon_n$ in the standard basis $(\epsilon_1,\ldots,\epsilon_n)$ of the $\gl(n)$ weight space.

\setcounter{equation}{0}
\section{Parafermions and the parity operator}
\label{sec3}

In this section we recall the approach of~\cite{SV2026} to introduce a parafermion parity operator.
A system of $n$ fermion creation and annihilation operators $a_i^\pm$ ($i=1,\ldots,n$) is determined by the anti-commutation relations
\begin{equation}
 \{ a_i^+,a_j^+\} = \{ a_i^-,a_j^-\} = 0, \quad \{ a_i^-,a_j^+\} = \delta_{ij}.
\label{aa}
\end{equation}
The fermion Fock space is generated by a vacuum vector $|0\rangle$ satisfying
$a_i^- |0\rangle =0$ for all $i$, $\langle 0 | 0\rangle = 1$ and the hermiticity conditions $(a_i^\pm)^\dagger = a_i^\mp$.
A set of orthonormal basis vectors of the Fock space is given by
\begin{equation}
|\theta\rangle = |\theta_1,\ldots,\theta_n\rangle= (a_1^+)^{\theta_1}\ldots (a_n^+)^{\theta_n} |0\rangle,
\quad \theta_i\in\{ 0,1\}.
\label{theta}
\end{equation}
On such a basis state the number operator $F=\sum_{i=1}^n a_i^+a_i^-$ has action
\begin{equation}
F |\theta\rangle = (\sum_{i=1}^n \theta_i)\, |\theta\rangle,
\end{equation}
and the so-called fermion parity operator $P$ takes the form $P=(-1)^F$. $P$ has eigenvalue $+1$ on states with an even number of fermions, 
and $-1$ on states with an odd number of fermions.
Moreover, for all indices $i\in\{1,\ldots,n\}$:
\begin{equation}
\{P,P\}=2,\qquad \{ P, a_i^\pm\}=0.
\label{ha}
\end{equation}

When the set of $2n+1$ operators $a_i^\pm$ ($i=1,\ldots,n$) and $P$ satisfy the quadratic relations~\eqref{aa} and~\eqref{ha}, 
then they also satisfy the following cubic or triple relations:
\begin{align}
& [[a_{i}^{\xi}, a_{j}^{\eta}], a_{k}^{\epsilon}]=
|\epsilon -\eta| \delta_{jk} a_{i}^{\xi} - |\epsilon -\xi| \delta_{ik}a_{j}^{\eta}, 
\label{a-rels} \\
& [[P,a_j^\eta],P]=-4 a_j^\eta, \label{ahh}\\
& [[a_i^\xi,P],a^\epsilon_k]= - |\epsilon-\xi| \delta_{ik} P, \label{aah}
\end{align}
where $i,j,k\in \{1,2,\ldots,n\}$ and $\eta, \epsilon, \xi \in\{+,-\}$ (to be interpreted as $+1$ and $-1$
in the algebraic expressions $\epsilon -\xi$ and $\epsilon -\eta$).

Following the ideas of Green~\cite{Green}, a {\em new system} of $2n+1$ generators $a_i^\pm$ ($i=1,\ldots,n$) and $P$ satisfying the triple relations~\eqref{a-rels}-\eqref{aah} (and  no longer required to satisfy~\eqref{aa} and~\eqref{ha}) was considered in~\cite{SV2026}. 
Clearly, the elements satisfying~\eqref{a-rels} are known as parafermion creation and annihilation operators~\cite{Green}.
The extra generator $P$ satisfying~\eqref{ahh}-\eqref{aah} is called the parafermion parity operator.

It is known since a long time that the Lie algebra generated by the elements $a_i^\pm$ ($i=1,\ldots,n$) subject to the triple relations~\eqref{a-rels} is the simple Lie algebra $\so(2n+1)=B_n$~\cite{KT, Ryan}.
The algebraic structure determined by the $2n+1$ generators $a_i^\pm$ ($i=1,\ldots,n$) and $P$ subject to the relations~\eqref{a-rels}-\eqref{aah} was determined in~\cite{SV2026}, and is the Lie algebra $\so(2n+2)$.
The proof of this property uses the following matrix representation of $\so(2n+2)$.
Let $B$ be the symmetric $(2n+2)\times(2n+2)$ matrix:
\begin{equation}
B=\left( \begin{array}{cccc} 0 & I_n & 0 & 0\\ I_n & 0 & 0 &0\\ 0 & 0 & 0 & 1\\ 0 & 0 & 1 & 0 \end{array} \right),
\end{equation}
where $I_n$ is the identity matrix of size $n\times n$.
The Lie algebra $\so(2n+2)$ consists of the matrices $X$ satisfying
\begin{equation}
X^t B + B X=0,
\end{equation}
where $X^t$ is the notation for the transpose of $X$.
Explicitly, these matrices are of the following form
\begin{equation}
X=\left( \begin{array}{cccc} a & b & x & y \\ d & -a^t & x' & y' \\ -{y'}^t & -y^t & f & 0 \\ -{x'}^t & -x^t & 0 & -f \end{array} \right),
\end{equation}
where $a$ is a $n\times n$ matrix, $b$ and $d$ are antisymmetric $n\times n$ matrices, $x,y,x',y'$ are $n\times 1$ matrices, and $f$ is a number.
Using the common unit matrices $e_{i,j}$ (consisting of all zeros, except a 1 at position $(i,j)$), a basis of the Cartan subalgebra $\h$ of $\so(2n+2)$ is given by the elements
\begin{equation}
h_i=e_{i,i}-e_{n+i,n+i}\quad (1\leq i\leq n), \quad h_{n+1}=e_{2n+1,2n+1}-e_{2n+2,2n+2}.
\end{equation}
The basis of $\h^*$ dual to $\{h_1, h_2, \ldots, h_{n+1}\}$ is denoted by $\{\epsilon_1,\epsilon_2,\ldots,\epsilon_{n+1}\}$.
The matrix form of the $2n+1$ generators is
\begin{align}
a^+_i&= e_{i,2n+1}-e_{2n+2,n+i}+e_{i,2n+2}-e_{2n+1,n+i} , \nonumber\\ 
a^-_i&= e_{2n+1,i}-e_{n+i,2n+2}+e_{2n+2,i}-e_{n+i,2n+1} , \label{matrixform1}\\  
P&=2 h_{n+1} = 2e_{2n+1,2n+1}-2e_{2n+2,2n+2}. \nonumber
\end{align}
These matrices satisfy the triple relations~\eqref{a-rels}, \eqref{ahh} and~\eqref{aah}. 

It should be emphasized that the generators $a_i^\xi$ of $\so(2n+2)$ are {\em not} root vectors of $\so(2n+2)$, but linear combinations of two root vectors.
The parity operator $P$ is just an element of the Cartan subalgebra.

\setcounter{equation}{0}
\section{The parafermion Fock spaces, characters and the parity operator}
\label{sec4}

The parafermion Fock space $W(p)$ (with $p$ a positive integer) 
is the Hilbert space with unique vacuum vector $|0\rangle$, 
defined by means of ($j,k=1,2,\ldots,n$)~\cite{Green, GM}
\begin{align}
& \langle 0|0\rangle=1, \qquad a_j^- |0\rangle = 0, \qquad (a_j^\pm)^\dagger = a_j^\mp,\nonumber\\
& [a_j^-,a_k^+] |0\rangle = p\,\delta_{jk}\,|0\rangle,
\label{pFock}
\end{align}
and by irreducibility under the action of the algebra generated by
the elements $a_j^+$ and $a_j^-$ ($j=1,\ldots,n$), subject to the triple relations~\eqref{a-rels}.
The parameter $p$ is the so-called order of statistics of the parafermion system.
The parafermion Fock space $W(p)$ is the unitary irreducible representation of
$\so(2n+1)$ with lowest weight $(-\frac{p}{2}, -\frac{p}{2},\ldots, -\frac{p}{2})$~\cite{parafermion}.
The highest weight of $W(p)$ is $(\frac{p}{2}, \frac{p}{2},\ldots, \frac{p}{2})=\frac{p}{2}\sum_{i=1}^n \epsilon_i$, with Dynkin labels $[0,\ldots,0,p]$.
A Gelfand-Zetlin basis for $W(p)$ was determined in~\cite{parafermion}.
It consists of the vectors of the following shape:
\begin{equation}
 |m)\equiv |m)^n\equiv \left|
\begin{array}{lcllll}
 m_{1n} & \cdots & \cdots & m_{n-1,n} & m_{nn}  \\
 m_{1,n-1} & \cdots & \cdots &  m_{n-1,n-1}  &  \\
\vdots & \qdots & & & \\
m_{11} & & & &
\end{array}
\right) 
= \left| \begin{array}{l} [m]^n \\[2mm] |m)^{n-1} \end{array} \right) \;.
\label{mn}
\end{equation}
The top row of this pattern, also denoted by the $n$-tuple $[m]^n$, is any
partition $\mu$, so we identify $\mu=(m_{1n}, m_{2n},\ldots, m_{nn})$. 
The remaining $n-1$ rows of the pattern will sometimes be denoted by $|m)^{n-1}$.
All $m_{ij}$ in the above GZ-pattern are non-negative integers,
satisfying the {\em betweenness conditions}
\begin{equation}
m_{i,j+1}\geq m_{ij}\geq m_{i+1,j+1}\qquad (1\leq i\leq j\leq n-1).
\label{between}
\end{equation}
In other words, two consecutive rows in~\eqref{mn} are interlaced partitions.
The GZ-basis of $W(p)$ consists of all vectors of the form~\eqref{mn} satisfying~\eqref{between} and $m_{1n}\leq p$. 
In other words, the top line of $|m)$ is a
partition $\mu$ with largest part not exceeding $p$: $\mu_1\leq p$.

How can this $\so(2n+1)$ representation $W(p)$ be extended to a representation of $\so(2n+2)$?
This extension is necessary in order to include the parity operator.
It is known that there are two representations of $\so(2n+2)$ that decompose into the single $\so(2n+1)$ representation $W(p)$, in the branching $\so(2n+2)\supset\so(2n+1)$~\cite{Okada}.
These representations will be denoted by $W_+(p)$ and $W_-(p)$, and the corresponding Dynkin labels are given in the following diagrams:
\[
\begin{array}{cc}
W_-(p) & W_+(p) \\
\qquad\begin{tikzpicture}[scale=0.5,
    dnode/.style={circle, draw=black, thick, minimum size=6pt, inner sep=0pt},
    sedge/.style={-, thick}]

    \node[dnode, label=below:$0$] (1) at (0,0) {};
    \node[dnode, label=below:$0$] (2) at (1,0) {};
    \node[dnode, label=below:$0$] (3) at (2.5,0) {};
    \node[dnode, label=below:$0$] (4) at (3.5,0) {};

    \path (1) edge[sedge] (2)
          (3) edge[sedge] (4);
          
		\draw[dashed] (2) -- (3);

    \node[dnode, label=above:$p$] (5) at (4.5,0.5) {};
    \node[dnode, label=below:$0$] (6) at (4.5,-0.5) {};

    \path (4) edge[sedge] (5)
          (4) edge[sedge] (6);

\end{tikzpicture} \qquad
&
\qquad\begin{tikzpicture}[scale=0.5,
    dnode/.style={circle, draw=black, thick, minimum size=6pt, inner sep=0pt},
    sedge/.style={-, thick}]

    \node[dnode, label=below:$0$] (1) at (0,0) {};
    \node[dnode, label=below:$0$] (2) at (1,0) {};
    \node[dnode, label=below:$0$] (3) at (2.5,0) {};
    \node[dnode, label=below:$0$] (4) at (3.5,0) {};

    \path (1) edge[sedge] (2)
          (3) edge[sedge] (4);
          
		\draw[dashed] (2) -- (3);

    \node[dnode, label=above:$0$] (5) at (4.5,0.5) {};
    \node[dnode, label=below:$p$] (6) at (4.5,-0.5) {};

    \path (4) edge[sedge] (5)
          (4) edge[sedge] (6);

\end{tikzpicture} \qquad
\end{array}
\]
Both of these decompose to $W(p)$, with the following labels in the $\so(2n+1)$ Dynkin diagram:
\[
\begin{tikzpicture}[scale=.4]
    \draw (-1,0) node[anchor=east]  {$W(p)$};
    \foreach \x in {0,...,3}
    \draw[xshift=\x cm,thick] (\x cm,0) circle (.3cm);
    \draw[xshift=4 cm,thick] (4 cm, 0) circle (.3 cm);
    \draw[thick] (0.3 cm,0) -- +(1.4 cm,0);
		\draw[thick,dashed] (2.3 cm,0) -- +(1.4 cm,0);
		\draw[thick] (4.3 cm,0) -- +(1.4 cm,0);
    \draw[thick] (6.3 cm, .1 cm) -- +(1.4 cm,0);
    \draw[thick] (6.3 cm, -.1 cm) -- +(1.4 cm,0);
		\draw[thick] (6.8 cm,0.3 cm) -- (7.1 cm, 0 cm) -- (6.8 cm, -0.3 cm); 
		\draw (0 cm,1 cm) node {$0$}; \draw (2 cm,1 cm) node {$0$}; \draw (4 cm,1 cm) node {$0$}; \draw (6 cm,1 cm) node {$0$}; \draw (8 cm,1 cm) node {$p$};
  \end{tikzpicture}
\]
So $W_+(p)$ is the $\so(2n+2)$ representation with Dynkin labels $[0,\ldots,0,p]$, or highest weight
$(\frac{p}{2}, \ldots,\frac{p}{2}, \frac{p}{2}) = \sum_{i=1}^{n+1} \frac{p}{2}\epsilon_i$,
and $W_-(p)$ is $\so(2n+2)$ representation $W_-(p)$ with Dynkin labels $[0,\ldots,p,0]$, or highest weight
$(\frac{p}{2}, \ldots, \frac{p}{2},-\frac{p}{2}) = \sum_{i=1}^{n} \frac{p}{2}\epsilon_i-\frac{p}{2}\epsilon_{n+1}$.
Both decompose to the single $\so(2n+1)$ representation $W(p)$ with highest weight $(\frac{p}{2}, \ldots, \frac{p}{2}) = \sum_{i=1}^n \frac{p}{2}\epsilon_i$ or 
Dynkin labels $[0,\ldots,0,p]$.

Since the representation space $W_\pm(p)$ of $\so(2n+2)$ is the same as the representation space $W(p)$ of $\so(2n+1)$,
we can keep the same notation $|m)$ for the basis vectors of $W_+(p)$ and $W_-(p)$.

Special formulas for the characters of $W_\pm(p)$ were determined in~\cite{Okada}.
As common, we put $x_i=e^{\epsilon_i}$ to write these characters.
First of all, using Theorem 2.2~(1) with $m=0$ or Lemma~5.4 of Ref.~\cite{Okada}, we have 
\begin{equation}
 \ch W_\pm(p)(x_1,\ldots,x_n,1) = \ch W(p)(x_1,\ldots,x_n).
\end{equation}
So this implies indeed the decomposition described above.

Next, let us apply Theorem~2.3~(3) of Ref.~\cite{Okada}, first in the case that $n$ is odd:
\begin{equation}
\ch W_+(p)(x_1,\ldots,x_n,x_{n+1})=(x_1\cdots x_{n+1})^{-p/2} \sum_{\lambda\in {\cal E}'; \lambda_1\leq p} s_\lambda(x_1,\ldots,x_n,x_{n+1}).
\end{equation}
Herein, ${\cal E}'$ is the set of partitions with $\lambda_{2i-1}=\lambda_{2i}$ for each $i\geq 1$, and $s_\lambda$ is the Schur polynomial.

It is well known~\cite{Macdonald} that in general
\begin{equation}
s_\lambda(x_1,\ldots,x_n,x_{n+1})=\sum_\mu s_\mu(x_1,\ldots,x_n) x_{n+1}^{|\lambda|-|\mu|},
\end{equation}
where the sum runs over all partitions $\mu$ such that $\lambda-\mu$ is a horizontal strip.

It is easy to see that if we remove a horizontal strip from a partition $\lambda$ of the set ${\cal E}'$, then $|\lambda|-|\mu|= |\mu|_a$. 
For example, when $\lambda=(7,7,4,4)$ and we remove a horizontal strip such that $\mu=(7,5,4,1)$, 
\[
\begin{Young}
&&&&&&\cr
&&&&&$\times$&$\times$\cr
&&&\cr
&$\times$&$\times$&$\times$\cr
\end{Young}
\]
then
\[
|\lambda|-|\mu| = 5 = |\mu|_a=7-5+4-1.
\]
More general, if $\lambda\in{\cal E}'$ and $\theta=\lambda-\mu$ is a horizontal strip consisting of $|\theta|=|\lambda|-|\mu|$ boxes,
then the number of boxes of $\theta$ is equal to the number of columns of $\mu$ with odd length, 
and following~\eqref{oddparts} this is indeed equal to $|\mu|_a$.

This implies
\begin{equation}
\ch W_+(p)(x_1,\ldots,x_n,x_{n+1})=(x_1\cdots x_{n+1})^{-p/2} \sum_{\mu_1\leq p} s_\mu(x_1,\ldots,x_n) x_{n+1}^{|\mu|_a},
\end{equation}
where the sum is over all partitions $\mu$ with $\mu_1\leq p$. 

Thus the eigenvalue of $h_{n+1}$ on a vector $|m)$ of $W_+(p)$, of which the top row equals the partition $\mu$, is given by
$-p/2+|\mu|_a$, and following~\eqref{matrixform1} this implies
\[
P |m) = (-p+2|\mu|_a) \; |m).
\]

Now let us apply Theorem~2.3~(3) of \cite{Okada} in the case that $n$ is even.
\begin{equation}
\ch W_+(p)(x_1,\ldots,x_n,x_{n+1})=(x_1\cdots x_{n+1})^{-p/2} \sum_{\lambda\in {\cal O}'} s_\lambda(x_1,\ldots,x_n,x_{n+1}).
\end{equation}
Herein, ${\cal O}'$ is the set of partitions with $\lambda_1=p$ and $\lambda_{2i}=\lambda_{2i+1}$ for each $i\geq 1$.
In this case, if we remove a horizontal strip from a partition of ${\cal O}'$, then $|\lambda|-|\mu|=p- |\mu|_a$. 
This implies
\begin{equation}
\ch W_+(p)(x_1,\ldots,x_n,x_{n+1})=(x_1\cdots x_{n+1})^{-p/2} \sum_{\mu_1\leq p} s_\mu(x_1,\ldots,x_n) x_{n+1}^{p-|\mu|_a},
\end{equation}
where the sum is over all partitions $\mu$ with $\mu_1\leq p$. 
Then the eigenvalue of $h_{n+1}$ on a vector $|m)$ of $W_+(p)$ with the top row the partition $\mu$, is given by
$p/2-|\mu|_a$, and this implies
\[
P |m) = (p-2|\mu|_a) \; |m).
\]

Combining the two cases ($n$ even and odd) we find:
\begin{equation}
P |m) = (-1)^n (p-2|\mu|_a) \; |m).
\label{P+-eigenvalue}
\end{equation}
This formula coincides with Eq.~(3.6) of Ref.~\cite{SV2026}. 
However, now the deduction of this formula is much simpler, compared to the complicated computations of~\cite{SV2026}.

For $W_-(p)$, the analysis is essentially the same, and we find
\begin{equation}
P |m) = -(-1)^n (p-2|\mu|_a) \; |m)
\label{P--eigenvalue}
\end{equation}
in agreement with~\cite[Eq.~(3.7)]{SV2026}.

\setcounter{equation}{0}
\section{Parabosons and the parity operator}
\label{sec5}

In this section, the approach of~\cite{SV2026} to the paraboson parity operator is recalled .
A system of $n$ boson creation and annihilation operators $a_i^\pm$ ($i=1,\ldots,n$) is determined by the commutation relations
\begin{equation}
 [ a_i^+,a_j^+] = [ a_i^-,a_j^-] = 0, \quad [ a_i^-,a_j^+] = \delta_{ij}.
\label{1-aa}
\end{equation}
The boson Fock space is generated by a vacuum vector $|0\rangle$ satisfying
$a_i^- |0\rangle =0$ for all $i$, $\langle 0 | 0\rangle = 1$,
and the hermiticity conditions $(a_i^\pm)^\dagger = a_i^\mp$.
A set of orthonormal basis vectors of the Fock space is given by
\begin{equation}
|k\rangle = |k_1,\ldots,k_n\rangle= \frac{1}{\sqrt{k_1!\cdots k_n!}}(a_1^+)^{k_1}\ldots (a_n^+)^{k_n} |0\rangle,
\quad k_i\in\N=\{ 0,1,2\ldots \}.
\label{k}
\end{equation}
The number operator is $F=\sum_{i=1}^n a_i^+a_i^-$, and its action in the Fock space is
\begin{equation}
F |k\rangle = (\sum_{i=1}^n k_i)\, |k\rangle.
\end{equation}
The boson parity operator $P$ takes the form $P=(-1)^F$,
and has eigenvalue $+1$ on states with an even number of bosons, 
and $-1$ on states with an odd number of bosons.
It satisfies
\begin{equation}
\{P,P\}=2,\qquad \{ P, a_i^\pm \} =0\qquad (i\in \{1,\ldots,n\}).
\label{1-ha}
\end{equation}

When the set of $2n+1$ operators $a_i^\pm$ ($i=1,\ldots,n$) and $P$ satisfy the quadratic relations~\eqref{1-aa} and~\eqref{1-ha}, it is easy to verify that they also satisfy the following triple relations:
\begin{align}
& [\{a_{i}^{\xi}, a_{j}^{\eta}\}, a_{k}^{\epsilon}]=
(\epsilon -\eta) \delta_{jk} a_{i}^{\xi} + (\epsilon -\xi) \delta_{ik}a_{j}^{\eta}, 
\label{1-a-rels} \\
& [[P,a_j^\eta],P]=-4a_j^\eta, \label{1-ahh}\\
& \{[a_i^\xi,P],a^\epsilon_k\}= - (\epsilon-\xi) \delta_{ik} P, \label{1-aah}
\end{align}
where $i,j,k\in \{1,2,\ldots,n\}$ and $\eta, \epsilon, \xi \in\{+,-\}$.

Following again the approach of Green~\cite{Green}, 
let us consider a {\em new system} of $2n+1$ generators $a_i^\pm$ ($i=1,\ldots,n$) and $P$ 
that are required to satisfy the triple relations~\eqref{1-a-rels}-\eqref{1-aah}.
The elements satisfying~\eqref{1-a-rels} are known as paraboson creation and annihilation operators~\cite{Green}.
The element $P$ satisfying~\eqref{1-ahh}-\eqref{1-aah} is the paraboson parity operator of~\cite{SV2026}.

The algebra generated by the elements $a_i^\pm$ ($i=1,\ldots,n$) subject to the triple relations~\eqref{1-a-rels} 
is known to be the Lie superalgebra $\osp(1|2n)$ (where $a_i^\pm$ are odd elements)~\cite{Ganchev}.
In~\cite{SV2026}, the algebra generated by the $2n+1$ generators $a_i^\pm$ ($i=1,\ldots,n$) and $P$ subject to the relations~\eqref{1-a-rels}-\eqref{1-aah} 
was identified as the Lie superalgebra $\osp(2|2n)$ (where $P$ is an even element).
In the proof of this property, the following matrix representation of $\osp(2|2n)$ is used~\cite{Kac}:
\begin{equation}
X=\left( \begin{array}{cccc} f & 0 & x & x' \\ 0 & -f & y & y' \\ {y'}^t & {x'}^t & a & b \\ -{y}^t & -x^t & c & -a^t \end{array} \right),
\end{equation}
where $a$ is a $n\times n$ matrix, $b$ and $c$ are symmetric $n\times n$ matrices, $x,y,x',y'$ are $1\times n$ matrices, and $f$ is a number.
The even matrices have $x=x'=y=y'=0$, and the odd matrices have $a=b=c=0$ and $f=0$.
Using the common unit matrices $e_{i,j}$, a basis of the Cartan subalgebra $\h$ of $\osp(2|2n)$ is given by the elements
\begin{equation}
h_{0}=e_{1,1}-e_{2,2}, \quad h_i=e_{2+i,2+i}-e_{2+n+i,2+n+i}\quad (1\leq i\leq n).
\end{equation}
The basis of $\h^*$ dual to $\{h_0, h_1, \ldots, h_{n}\}$ is denoted by $\{\epsilon_0,\epsilon_1,\ldots,\epsilon_{n}\}$.
Now the matrix form of the $2n+1$ generating elements is
\begin{align}
a_i^+ &= -e_{1,2+n+i}-e_{2+i,2}+e_{2,2+n+i}+e_{2+i,1}, \nonumber\\
a_i^- &=  e_{2+n+i,1}-e_{2,2+i}-e_{2+n+i,2}+e_{1,2+i}, \\
P &= 2h_0 = 2 e_{1,1}-2 e_{2,2}. \nonumber
\end{align}
The fact that these matrices satisfy the triple relations~\eqref{1-a-rels}-\eqref{1-aah} is used in~\cite{SV2026} to prove that the underlying algebra is indeed $\osp(2|2n)$.

\setcounter{equation}{0}
\section{The paraboson Fock space, the parity operator and $\osp(2|2n)$ characters}
\label{sec6}

The paraboson Fock space $V(p)$ is the Hilbert space with unique vacuum vector $|0\rangle$, 
defined by means of ($j,k=1,2,\ldots,n$)~\cite{Green, GM}
\begin{align}
& \langle 0|0\rangle=1, \qquad a_j^- |0\rangle = 0, \qquad (a_j^\pm)^\dagger = a_j^\mp,\nonumber\\
& \{a_j^-,a_k^+\} |0\rangle = p\,\delta_{jk}\,|0\rangle,
\label{1-pFock}
\end{align}
and by irreducibility under the action of the algebra generated by
the elements $a_j^+$ and $a_j^-$ ($j=1,\ldots,n$), subject to the triple relations~\eqref{1-a-rels}.
The parameter $p$ (a positive integer) is the order of statistics of the paraboson system.
It is known that the paraboson Fock space $V(p)$ is the unitary irreducible representation of
$\osp(1|2n)$ with lowest weight $(\frac{p}{2}, \frac{p}{2},\ldots, \frac{p}{2})$.
A Gelfand-Zetlin basis for $V(p)$ was determined in~\cite{paraboson}, and consists of the vectors $|m)$ of the same form as in~\eqref{mn}.
The top row of $|m)$ is again a partition $\mu$, and consecutive rows in~\eqref{mn} are interlaced partitions.
The GZ-basis of $V(p)$ consists of all vectors of the form~\eqref{mn} satisfying $\ell(\mu)\leq p$;
in other words: $\mu$ is a partition with at most $p$ nonzero parts (and there is no restriction on the parts themselves).
So if $p\geq n$, there is no extra restriction on $\mu$ (since the top row of~\eqref{mn} always has length at most $n$).
$V(p)$ is an infinite-dimensional representation since there is no restriction on the parts of $\mu$.

In~\cite{SV2026} we showed that there are two irreducible representations of $\osp(2|2n)$, denoted by $V_+(p)$ and $V_-(p)$, whose representation spaces
coincide with $V(p)$, and thus their basis can also be denoted by $|m)$. Moreover, the action of the generators $a_i^\pm$ on $|m)$ is the same as for $V(p)$. 
The action of the extra generator $P$ was given in~\cite[(5.9)-(5.10)]{SV2026}.
Using formula~\eqref{26}, this action can be written as follows:
\begin{align}
\hbox{in } V_+(p):\quad & P|m) = \left(p-2|\mu'|_a\right) \ |m), \nonumber \\
\hbox{in } V_-(p):\quad & P|m) = - \left(p-2|\mu'|_a\right) \ |m). 
\label{P-action}
\end{align}
Note that the spectrum of $P$ is equal to $\{p,p-2,\ldots,-p\}$ when $p\leq n$ and to $\{p,p-2,\ldots,p-2n\}$ when $p\geq n$.

Let us describe the representations appearing here in terms of their lowest weight.
For $\osp(2|2n)$ we work with the following Dynkin diagram and set of simple roots~\cite{Kac,Frappat}:
\[
\begin{tikzpicture}[scale=.8]
    \draw (-1,0) node[anchor=east]  {$\osp(2|2n)$};
		\draw[thick,fill=lightgray] (0 cm,0) circle (0.3 cm); 
    \foreach \x in {1,...,3}
    \draw[xshift=\x cm,thick] (\x cm,0) circle (.3cm);
    \draw[xshift=4 cm,thick] (4 cm, 0) circle (.3 cm);
    \draw[thick] (0.3 cm,0) -- +(1.4 cm,0);
		\draw[thick,dashed] (2.3 cm,0) -- +(1.4 cm,0);
		\draw[thick] (4.3 cm,0) -- +(1.4 cm,0);
    \draw[thick] (6.3 cm, .1 cm) -- +(1.4 cm,0);
    \draw[thick] (6.3 cm, -.1 cm) -- +(1.4 cm,0);
		\draw[thick] (7.1 cm,0.3 cm) -- (6.8 cm, 0 cm) -- (7.1 cm, -0.3 cm); 
		\draw (0 cm,0.6 cm) node {$\epsilon_0-\epsilon_1$}; \draw (2 cm,0.6 cm) node {$\epsilon_1-\epsilon_2$}; \draw (4 cm,0.6 cm) node {$\ldots$}; \draw (6 cm,0.6 cm) node {$\epsilon_{n-1}-\epsilon_n$}; \draw (8 cm,0.6 cm) node {$2\epsilon_n$};
  \end{tikzpicture}
\]
In this notation, $V_+(p)$ is the irreducible lowest weight representation with lowest weight given by 
\[
(\frac{p}{2}, \frac{p}{2}, \ldots,\frac{p}{2})=\frac{p}{2}\epsilon_0+\sum_{i=1}^n \frac{p}{2}\epsilon_i,
\]
and $V_-(p)$ is the irreducible lowest weight representation with lowest weight given by 
\[
 (-\frac{p}{2}, \frac{p}{2}, \ldots,\frac{p}{2})=-\frac{p}{2}\epsilon_0+\sum_{i=1}^n \frac{p}{2}\epsilon_i.
\]
For $\osp(1|2n)$, we work with the following Dynkin diagram and simple roots:
\[
\begin{tikzpicture}[scale=.8]
    \draw (-1,0) node[anchor=east]  {$\osp(1|2n)$};
    \foreach \x in {0,...,3}
    \draw[xshift=\x cm,thick] (\x cm,0) circle (.3cm);
    \draw[xshift=4 cm,thick,fill=black] (4 cm, 0) circle (.3 cm);
    \draw[thick] (0.3 cm,0) -- +(1.4 cm,0);
		\draw[thick,dashed] (2.3 cm,0) -- +(1.4 cm,0);
		\draw[thick] (4.3 cm,0) -- +(1.4 cm,0);
    \draw[thick] (6.3 cm, .1 cm) -- +(1.4 cm,0);
    \draw[thick] (6.3 cm, -.1 cm) -- +(1.4 cm,0);
		\draw[thick] (6.8 cm,0.3 cm) -- (7.1 cm, 0 cm) -- (6.8 cm, -0.3 cm); 
		\draw (0 cm,0.6 cm) node {$\epsilon_1-\epsilon_2$}; \draw (2 cm,0.6 cm) node {$\epsilon_1-\epsilon_2$}; \draw (4 cm,0.6 cm) node {$\ldots$}; \draw (6 cm,0.6 cm) node {$\epsilon_{n-1}-\epsilon_n$}; \draw (8 cm,0.6 cm) node {$\epsilon_n$};
  \end{tikzpicture}
\]
The paraboson Fock space $V(p)$ is the irreducible lowest weight representation with lowest weight
\[
 (\frac{p}{2}, \frac{p}{2}, \ldots,\frac{p}{2})=\sum_{i=1}^n \frac{p}{2}\epsilon_i.
\]
So our result is that in the branching $\osp(2|2n) \supset \osp(1|2n)$, both $V_+(p)$ and $V_-(p)$ decompose to the single representations $V(p)$.

For the case of parafermions and $\so(2n+2)\supset\so(2n+1)$, we could work with the known character formulas to deduce the action of $P$.
In the present case of parabosons, characters of infinite-dimensional representations (like $V_+(p)$ and $V_-(p)$) are not known in general.
But using the computational result of~\cite{SV2026}, given by~\eqref{P-action}, we can work the other way round.
Indeed, the character of the $\osp(1|2n)$ representation $V(p)$ is given by~\cite{paraboson}
\[
\ch V(p)(x_1,\ldots,x_n)=(x_1\cdots x_{n})^{p/2} \sum_{\ell(\mu)\leq p} s_\mu(x_1,\ldots,x_n),
\]
where $x_i=e^{\epsilon_i}$ and $s_\mu$ is the Schur function.
But now the action of the extra Cartan element $h_0=P/2$ in $\osp(2|2n)$ is known from~\eqref{P-action}.
Hence we can write:
\begin{align}
&\ch V_+(p)({x_0},x_1,\ldots,x_n)=({x_0}x_1\cdots x_{n})^{p/2} \sum_{\ell(\mu)\leq p} {x_0^{-|\mu'|_a}}s_\mu(x_1,\ldots,x_n), \label{charV+}\\
&\ch V_-(p)({x_0},x_1,\ldots,x_n)=({x_0^{-1}}x_1\cdots x_{n})^{p/2} \sum_{\ell(\mu)\leq p} {x_0^{|\mu'|_a}}s_\mu(x_1,\ldots,x_n). \label{charV-}
\end{align}
As far as we know, these are new characters for $\osp(2|2n)$ representations.
The nice feature is that the alternating sum of parts of partitions appears naturally in these character formulas.

Once such a character formula is known, one can deduce other characteristics, for example generating functions.
Let us consider the generating function for the $P$-eigenvalues in $V_+(p)$.
Following~\eqref{charV+}, this is
\begin{equation}
\sum_{\mu,\; \ell(\mu)\leq p} {\cal P}^{p-2|\mu|_a} \dim(\mu) q^\mu =
\sum_{\mu,\; \ell(\mu)\leq p} {\cal P}^{p-n+(-1)^{\mu_1}+\cdots +(-1)^{\mu_n}} \dim(\mu) q^\mu.
\label{GF}
\end{equation}
Herein, $q^\mu$ is a short hand notation for $q_1^{\mu_1}\cdots q_n^{\mu_n}$ and the dummy variables $q_1,\ldots,q_n$ keep track of the $n$-tuple $\mu$,
which is in fact the highest weight of the $\gl(n)$ representation with highest weight $\mu$ in the decomposition of $V_+(p)$ according to $\osp(2|2n)\supset\osp(1|2n)\supset\gl(n)$.
Thus $\dim(\mu)$ stands for the dimension of this $\gl(n)$ representation and is given by
\[
\dim(\mu)= \prod_{1\leq i<j\leq n} \frac{\mu_i-\mu_j+j-i}{j-i}.
\]

Let us consider an example for $n=3$, i.e.\ the $3D$ paraboson oscillator.
For $p=1$, \eqref{GF} yields (terms up to degree 5 for $|\mu|$):
\[
{\cal P}+ 3{\cal P}^{-1}q_1+ 6{\cal P}q_1^2+ 10{\cal P}^{-1}q_1^3+ 15{\cal P}q_1^4+ 21{\cal P}^{-1}q_1^5 + \cdots
\]
Since for $p=1$ the parabosons are just ordinary bosons, this is the parity operator generating function for the $3D$ boson oscillator in the Fock space.
The eigenvalues of $P$ are indeed $+1$ and $-1$ (powers of ${\cal P}$); 
the powers of $q_i$ keep track of the $\gl(3)$ representations, which are in this case just the symmetric representations of the form $(\mu_1,0,0)$;
and the numerical coefficients in front of each term correspond to the dimensions of these representations.

For $p=2$, \eqref{GF} yields (again terms up to degree 5 for $|\mu|$):
\begin{align*}
& {\cal P}^2+ 3q_1+ 6{\cal P}^2q_1^2+ 3{\cal P}^{-2}q_1q_2+ 10q_1^3+ 8q_1^2 q_2\\
&+ 15{\cal P}^2q_1^4+ 15{\cal P}^{-2}q_1^3q_2+ 6{\cal P}^2q_1^2q_2^2 + 21q_1^5+ 24q_1^4q_2+15q_1^3q_2^2+\ldots,
\end{align*}
with the spectrum of $P$ being $\{-2,0,2\}$.

For $p=3$ or $p\geq 3$, one finds
\begin{align*}
&{\cal P}^p+ 3 {\cal P}^{p-2} q_1+ 6 {\cal P}^p q_1^2+ 3 {\cal P}^{p-4}q_1 q_2 + 10 {\cal P}^{p-2} q_1^3+ 8 {\cal P}^{p-2} q_1^2 q_2+ {\cal P}^{p-6}q_1 q_2 q_3\\
 &+15 {\cal P}^p q_1^4+ 15{\cal P}^{p-4} q_1^3 q_2+ 6 {\cal P}^p q_1^2 q_2^2+ 3{\cal P}^{p-4} q_1^2 q_2 q_3+ 21 {\cal P}^{p-2} q_1^5+ 24 {\cal P}^{p-2} q_1^4 q_2+ 15 {\cal P}^{p-2} q_1^3 q_2^2\\
& + 6{\cal P}^{p-6} q_1^3 q_2 q_3+ 3 {\cal P}^{p-2} q_1^2 q_2^2 q_3+\ldots
\end{align*}

\setcounter{equation}{0}
\section{Conclusion}
\label{sec7}
The introduction of a parity operator $P$ for a set of $n$ parafermions or parabosons has many interesting aspects.
In~\cite{SV2026}, where this operator $P$ was introduced for the first time, the algebraic properties were emphasized.
In particular, the underlying algebras $\so(2n+1)$ and $\osp(1|2n)$ (for parafermions and parabosons respectively) were shown to extend to $\so(2n+2)$ and $\osp(2|2n)$.
The action of $P$ in the relevant Fock spaces was also computed in~\cite{SV2026}, but it was not clear why the spectrum of $P$ turned out to be so simple.

In the current paper, we have reexamined this problem, using particular character formulas of the Fock spaces. 
For the parafermion Fock spaces $W(p)$, the character in the $\so(2n+2)$ representations $W_\pm(p)$ leads to a simple proof of the $P$-eigenvalues in terms of the alternating sum of parts $|\mu|_a$ of a partition $\mu$. 
This terminology turns out to be very significant, and we showed that for the paraboson Fock spaces $V(p)$ the $P$-eigenvalue takes the same form with $|\mu|_a$ replaced by $|\mu'|_a$.
This similarity is very striking, and strengthens the current combinatorial approach.

For the paraboson Fock spaces, this approach led to new character formulas for certain infinite-dimensional $\osp(2|2n)$ representations.

The central combinatorial notion in this paper is the alternating sum of parts of a partition.
This quantity has appeared in some papers, usually related to generalizations of Euler's partition theorem, see e.g.~\cite{Chen}. 
As far as we know, it is the first time it appears in character formulas, and in particular as the eigenvalue of an operator in a physical context.



\paragraph{Funding information}
Both authors were supported by the Bulgarian National Science Fund, Grant KP-06-N88/3.

\



\bibliographystyle{SciPost_bibstyle}



\end{document}